
\magnification\magstep1
\hsize 15.5truecm

\def\a{\alpha}\def\b{\beta}\def\d{\delta}

\def\la{\lambda}\def\La{\Lambda}

\def\f{\varphi}

\def\CC{{\bf C}}


\def\la{\lambda}

\def\sqr#1#2{{\vcenter{\hrule height.#2pt%
\hbox{\vrule width.#2pt height#1pt\kern#1pt%
\vrule width.#2pt}%
\hrule height.#2pt}}}

 at 17.28truept
 at 14.4truept
 at 12truept
 at 10.95truept
 at 17.28truept
 at 10truept

\def\title#1{%
\vskip0pt plus.3\vsize\penalty-100%
\vskip0pt plus-.3\vsize\bigskip\vskip\parskip%
\bigbreak\bigbreak\centerline{\bf #1}\bigskip%
}

\def\chapter#1#2{\vfill\eject
\centerline{\bf Chapter #1}
\vskip 6truept%
\centerline{\bf #2}%
\vskip 2 true cm}

\def\section#1#2{%
\def\\{#2}%
\vskip0pt plus.3\vsize\penalty-100%
\vskip0pt plus-.3\vsize\bigskip\vskip\parskip%
\par\noindent{\bf #1\hskip 6truept%
\ifx\empty\\{\relax}\else{\bf #2\smallskip}\fi}}

\def\subsection#1#2{%
\def\\{#2}%
\vskip0pt plus.3\vsize\penalty-20%
\vskip0pt plus-.3\vsize\medskip\vskip\parskip%
\def\TEST{#1}%
\noindent{\ifx\TEST\empty\relax\else\bf #1\hskip 6truept\fi%
\ifx\empty\\{\relax}\else{#2\smallskip}\fi}}

\def\proclaim#1{\medbreak\begingroup\noindent{\bf #1.---}\enspace\sl}

\def\endproclaim{\endgroup\par\medbreak}

\def\qqbox#1{\quad\hbox{#1}\quad}


\def\comfig#1#2\par{
\medskip
\centerline{\hbox{\hsize=10cm\eightpoint\baselineskip=10pt
\vbox{\noindent #1}}}\par\centerline{ Figure #2}}

\def\figcom#1#2\par{
\medskip
\centerline
{Figure #1}
\par\centerline{\hbox{\hsize=10cm\eightpoint\baselineskip=10pt
\vbox{\noindent #2}}}}

\def\adresse#1{%
\bigskip\hfill\hbox{\vbox{%
\hbox{#1}\hbox{Centre de Math\'ematiques}\hbox{Ecole Polytechnique}%
\hbox{F-91128 Palaiseau Cedex (France)}\hbox{\strut}%
\hbox{``U.A. au C.N.R.S. n$^{\circ}$169''}}}}


\def\comfig#1#2\par{
\medskip
\centerline{\hbox{\hsize=10cm\eightpoint\baselineskip=10pt
\vbox{\noindent{\sl  #1}}}}\par\centerline{{\bf Figure #2}}}

\def\figcom#1#2\par{
\medskip
\centerline
{{\bf Figure #1}}
\par\centerline{\hbox{\hsize=10cm\eightpoint\baselineskip=10pt
\vbox{\noindent{\sl  #2}}}}}

\def\em{\sl}

\def\\{\hfill\break}

\def\bibitem{\item}

\long\def\adresse#1{%
\leftskip=0truecm%
\vskip 3truecm%
\hbox{\hskip 10.5truecm{\hsize=7.5truecm\vbox{%
\def\cr{\par\noindent}\noindent#1}}}}


\title{Nilpotent action on the KdV variables and 2-dimensional Drinfeld-Sokolov
reduction}

\centerline{\rm B. Enriquez}

\medskip
{\bf Abstract.} {\em
We note that a version ``with spectral parameter'' of the
Drinfeld-Sokolov reduction gives a natural mapping from the KdV phase space to
the group of loops with values in $\widehat N_{+}/A, \widehat N_{+}$~: affine
nilpotent and $A$ principal commutative (or anisotropic Cartan) subgroup~; this
mapping is connected to the conserved densities of the hierarchy. We compute
the Feigin-Frenkel action of $\widehat n_{+}$ (defined in terms of screening
operators) on the conserved densities, in the $sl_2$ case.}

\medskip
\noindent
{\bf 1. 2-dimensional Drinfeld-Sokolov reduction.} Let $g = s\ell_{n}(\CC)$,
$b$ be
the Borel subalgebra of lower triangular matrices, $n$ be the nilpotent
subalgebra of strictly lower triangular matrices. Let $q(t)$ be a function of
$C^{\infty} (S^{1},b)$ and $I$ be the matrix $\sum^{n-1}_{i=1}e_{i,i+1}$,
$\La $
be the matrix $\sum^{n-1}_{i=1} e_{i,i+1} +\la e_{n,1}$ (the matrix elements
of $e_{ij}$ are $(e_{ij})_{\a\b} =\d_{ij,\a\b}$ for $1 \leq i,j,\a,\b \leq
n$). The according to [DS], the integrals of motion of the KdV equations,
associated to the Lax operator $\partial + I +q$, can be computed as follows~:
the operator $\partial + \La + q$ can be written in the form $n_{+} (x)
(\partial +\La +\sum^{\infty}_{i=0} u_{i}(x)\La^{-2}) n^{-1}_{+}(x)$, $n_{+}$,
$u_{i}$ differential polynomials in $q$, $u_{i}$ scalar and $n_{+}(x)$ matrix
valued in the Lie group associated to the Lie algebra $n + \la^{-1} g
[[\la^{-1}]]$.

Such a decomposition being fixed, all the other are of the form $(n_{+}(x)
\break
e^{
\sum^{\infty}_{i=0} v_{i} (x)\La^{-i}} , (u_{i}-v'_{i})(x))$, $v_{i}(x)$ being
arbitrary differential polynomials in $q(x)$.

Let $\widehat g$ be the Kac-Moody algebra $C^{\infty}(S^{1}, g)\oplus \CC$,
$\widehat n$ be its subalgebra $C^{\infty}(S^{1},n)$ and let us pose $\bar g =
\widehat g((\la^{-1})), \bar n = \widehat n + \la^{-1}\widehat g[[\la^{-1}]]$.
Then there is character $\bar p$ of $\bar n$ defined by $\bar p (f(t)
e_{ii+1}) =\int_{S^{1}} f$, $\bar p (f(t) \la^{-1} e_{n1})
= \int_{S^{1}} f$ and
$\bar p$ (all other $f(t)\la^{-k} e_{\a\b})=0$. Considering the projection
$\bar \pi : \bar g^{*} \to \bar n^{*}$ we can consider the Hamiltonian
reduction $\bar \pi^{-1} (\bar p)/ \bar N$. The resulting manifold will be
$\{ \partial + \La + \sum^{\infty}_{i=0} u_{i} (x)\La^{-i}\}/(u_{i}\sim u_{i} +
v'_{i})$ so that the coordinates on it are the polynomials in $\int_{S^{1}}
u_{i} (x)dx$, i.e. the integrals of motion. So the result of this reduction
is a Poisson commutative manifold. Giving ``Miura coordinates'' for it
equivalent to giving the expression of the $\int u_{i}$ in term of free fields.
It could be interesting to apply BRST techniques to quantize this reduction
(althrough the main result of these techniques, the interpretation of
the integrals of motion as intersection of kernels of screening operators, is
already known), for example we expect that for any $\bar g$-module $\bar M$ the
cohomology $H^{*}(\bar n ,\bar M)$ will be a module for the algebra of quantum
integrals of motion.

Notice that the result of [DS] also gives a natural map $\{ \partial + I +
q\}/\widehat N \to \widehat N\backslash \bar N / \bar A$, $\widehat N$ the Lie
group
associated to $\widehat n$ (it acts by conjugation on the left side), and
$\bar N$ and $\bar A$ the Lie groups corresponding to $\bar n$ and $\bar a =
C^{\infty} (S^1, \oplus_{\scriptstyle k \geq 0 \scriptstyle \atop n+k} \CC
\La^{-k}$). We will now study further this map.

\medskip
\noindent
{\bf 2. Nilpotent action on the KdV variables.}

\noindent
2.1. We first describe some recent results of B.~Feigin and E.~Frenkel [FF]. We
further restrict our selves to the case $g = s\ell_{2}(\CC)$.

Let $\phi$ be a free field on $S^{1}, \{ \phi (x),\phi(y)\}=\d'(x-y)$, and
consider the space of operators $\partial + I + \phi(x)h, h=\pmatrix{1 & 0\cr
0 & -1\cr}$. There are vector fields $Q_{+} = e^{-2\f} \{ \int_{S^{1}} e^{2\f},
\}$ and $Q_-= e^{2\f} \{ \int_{S^{1}} e^{-2\f},\}$ acting on this manifold and
generating an action of the Lie algebra $\widehat n_{+} = n_{+} \oplus
\nu^{-1} g [[ \nu^{-1}]]$. Here $n_{+}=\CC f , f = \pmatrix{0 & 0\cr 1 &
0\cr}$. Let $\pi_{x}$ be the set of functions on $\{ \partial + I + \phi(x)h\}$
of the form $P(\phi(x),\phi'(x)\cdots), P$, polynomial, then $\pi_{x}$ is a
$\widehat n_{+}$-module. In fact, $\pi_{x}\simeq (U \widehat n_{+}
\otimes_{Ua}\CC)^{*}$, $a = \Pi_{n\geq 0} \CC \bar \La^{-2n-1} (\bar\La=
\pmatrix{0 & 1\cr \nu & 0\cr})$, the pairing being
$$
\pi_{x} \times (U \widehat n_{+}\otimes_{U a}\CC) \to \CC
$$
$$
(P (\phi(x),\cdots ), T\otimes 1) \mapsto (TP (\phi))(\phi =0, \phi'=0,\cdots)=
\qqbox{constant term of} TP(\phi)\ .
$$
We thus may identify $\pi_{x}$ with a set of functions on $\widehat N_{+}/A$.

Feigin and Frenkel also claim that the KdV vector fields on $\pi_{x}$ may be
identified with the right infinitesimal action of $A_{-}$ on
$\widehat N_{+}/A$,
considered as a subspace of $\widehat B_{-} \setminus G((\nu^{-1}))/A$, $G
((\nu^{-1}))$ and $\widehat B_{-}$ being the Lie groups corresponding to the
Lie algebras $g((\la^{-1}))$ and
$\widehat b_{-}= b_{-} \oplus \nu g[\nu]$ ($b_{-}= \CC e \oplus \CC h$, $e =
\pmatrix{0 & 1\cr 0 & 0\cr}$, $h = \pmatrix{1 & 0\cr 0 & -1\cr})$, and
$A_{-}$ is the Lie group corresponding to the Lie algebra $a_{-} =
\oplus_{n\geq 0} \CC \bar \La^{2n+1}$. More precisely, the action of $\bar
\La^{2n+1} \in a_{-}$ is identified with the $n$-th KdV vector field (denoted
$\partial_{n})$. By the identification described above of $\pi_{x}$ with
the space of functions $\CC [N_{+}/A]$, the $n$-th KdV flow is
transformed into the vector field $R_{(n_{+}a_{-n} n^{-1}_{+})_{+}}$ at the
point $n_{+}A$ $(R$ is the infinitesimal right translation, $a_{-n} =
\bar\La^{2n+1}$, and the subscript $+$ denotes the projection on the first
factor
of the decomposition $g (( \la^{-1})) = \widehat n_{+}\oplus \widehat
b_{-})$. On the dual space $U\widehat n_{+} \otimes_{Ua}\CC$, this vector field
becomes the infinitesimal transformation
$\d_{a_{-n}}(T\otimes 1)= (T^{(1)} a_{-n} S(T^{(2)}))_{+} T^{(3)}\otimes 1$,
where we use Sweedler's notation for
coproducts, and $S$ denotes the antipode of $U\widehat n_{+}$. We note the
analogy of these formulas with those defining the Radul action.

\medskip
\noindent
2.2. From the considerations of 1 it is clear that if we pose $P_{+} = {1\over
2} (1+ {\La\over \sqrt{\la}})$, the matrix $n_{+}(x) P_{+} n_{+}(x)^{-1}$ will
be independent of the choice of $n_{+}(x)$, and a differential polynomial in
$\phi(x)$. We can see that this matrix is equal to
$$
M_{x} (\la) = {1\over 2\sqrt{\la}} \pmatrix{\psi_{\la}^{*}\cr -(\partial +\phi)
\psi^{*}_{\la}\cr} (x) (( \partial +\phi)\psi_{\la}\ \psi_{\la})(x)\ ,
$$
$\psi_{\la} , \psi^{*}_{\la}$ being conjugated Baker-Akhiezer functions for the
operator $L = (\partial -\phi )(\partial +\phi)$. Let $|1> = \pmatrix{1\cr0\cr}
, |2> = \pmatrix{0\cr 1\cr}$, and $\langle i|j\rangle =\d_{ij}$. Then $\langle
1|M_{x}(\la)|2 \rangle= {1\over 2\sqrt{\la}} \psi_{\la} \psi^{*}_{\la}(x)$. By
a statement of L.A.~Dickey [Di], ${\rm res}_{\sqrt{\la}=\infty}
(\sqrt{\la}^{n}\psi_{\la} (x)\psi^{*}_{\la}(x)d \sqrt{\la})={\rm
res}_{\partial} L^{n/2}$, so that $ \langle 1| M_{x} (\la)|2\rangle = {1\over
2\la} \sum_{n\geq -1} \la^{-{n\over 2}} {\rm res} L^{n/2}$. In the same way,
$\langle 2|M_{x}(\la)|1\rangle = {1\over
2}\sum_{n\geq -1} \la^{-{n\over 2}} {\rm res} \widetilde L^{{n\over 2}}$,
where $\tilde L = (\partial +\phi)(\partial -\phi)$. We will to compute the
action of $\widehat n_{+}$ on these elements.

First recall how to obtain the Lax operators of KdV from the mKdV one. We
have the conjugations
$$
\partial + \La + q_{0}= \partial + \La + \pmatrix{0 & 0\cr -\phi' +\phi^{2} &
0\cr} = \pmatrix{1 & 0\cr \phi & 1\cr} [ \partial + \La +
\pmatrix{\phi&0\cr 0&-\phi\cr} ] \pmatrix{1 & 0\cr \phi &
1\cr}^{-1}\ ,\leqno(1)
$$
and
$$
\partial + \La + q_{1}= \partial + \La +
\pmatrix{
0 &{\phi'+\phi^{2}\over\la}\cr
0 & 0\cr}
 =
\pmatrix{
1 & - {\phi\over\la}\cr 0 & 1\cr}
[ \partial + \La +
\pmatrix{
\phi &0\cr 0 &-\phi\cr} ]
\pmatrix{
1 &-{\phi \over\la}\cr
0 & 1\cr}^{-1}\ .\leqno(2)
$$
For $i=0,1,\quad \partial + \La + q_{i} = n_{i}n_{+}(x)(\partial +\La+
\sum_{i\geq 0} u_{i}(x)\La^{-i}) (n_{i}n_{+}(x))^{-1}$, with $n_{0}= \pmatrix{1
& 0\cr \phi & 1\cr}$, $n_{1}= \pmatrix{1 & - {\phi\over \la}\cr 0 & 1\cr}$. Let
us pose $M^{i}_{x}(\la) = n_{i}(x)M_{x}(\la)n_{i}(x)^{-1}$

It is then clear that $\langle 1|M_{x}(\la)| 2\rangle=\langle
1|M^{0}_{x}(\la)|2\rangle$, and $\langle 2 |M_{x}(\la)|1\rangle = \langle 2
|M^{1}_{x} (\la)| 1\rangle$.

Since $M^{0}_{x}(\la)$ (resp. $\widetilde M^{1}_{x}(\la))$ is a
differential polynomial in $q = - \phi'+\phi^{2}$ $({\rm resp.}\widetilde q
= \phi'+\phi^{2})$ the action of $Q_{+}\ ({\rm resp.}\ Q_{-})$ on it is
trivial.

Posing $M_{x}(\la)= \pmatrix{a & c\cr b & d\cr}$, we have then
$$
M^{0}_{x}(\la)= \pmatrix{a-c \phi & c\cr * & d+c\phi} , \, \widetilde
M^{1}_{x}(\la) = \pmatrix{a- {b\phi\over \la} & *\cr {b\over \la} & d +
{b\phi\over \la} \cr}.
$$
Recalling that $Q_{+}\phi =-2, Q_{-}\phi=2$, we deduce then
$$
Q_{+} a = - 2 c , Q_{+} c=0, Q_{+} d = 2 c \qqbox{and}
$$
$$
Q_{-} a = {2 b\over \la} , Q_{-} b=0 , Q_{-} d = - {2b\over \la} \ .
$$
The application of $Q_{+}\ ({\rm resp.}\  Q_{-})$ to the formula $a=a^{2}+bc$
gives
$Q_{+}b = 2 (a-d)\ ({\rm resp.}\ Q_{-} c = - {2\over \la} (a-d))$ (using
$a+d=1$).
So, we have $Q_{+}M_{x}(\la) = [ \pmatrix{0 & 0\cr 2 & 0\cr},\break
 M_{x}(\la)]$ and
$Q_{-}M_{x}(\la) = [ \pmatrix{0 & {2\over \la}\cr 0 & 0\cr}, M_{x}(\la)]$. We
deduce~:

\proclaim{Proposition}

1) The Feigin-Frenkel action of the element $T \in U \widehat n_{+}$ on the
matrix
$M_{x}(\la)$ (naturally extended by $\CC((\la^{-1/2}))$-linearity) is the
adjoint action
$$
\rho (T^{(1)}) M_{x}(\la)\rho (S({T^{(2)}}))\ ,
$$

where $\rho$ is the representation of $U\widehat n_{+}$ in
$SL_{2}(\CC((\la^{-1/2})))$ defined by $\rho(Q_{+})= \pmatrix{0 & 0\cr 2 &
0\cr}$,
$\rho (Q_{-})= \pmatrix{0 & {2\over \la}\cr 0 & 0\cr}$, $T^{(1)}
\otimes
T^{(2)}$ is the coproduct of $T$ and $S$ is the antipode of $U\widehat n_{+}$.

2) The pairing between $T$ and $M_{x}(\la)$ is equal to
$$
\rho (T^{(1)}) P_{+} \rho (S(T^{(2)}))\ ,\ P_{+} = {1\over 2} ( 1 + {\La\over
\sqrt{\la}} )\ .
$$
\endproclaim
\noindent
3) {\bf Bilinear relations.} Let us show how the bilinear relations
$\partial_{n} {\rm res} L^{m+ {1\over 2}}(x)= \partial_{m} {\rm res}
L^{n+{1\over 2}}(x)$ (recall that $\partial_{n}$ are the KdV flows) can be
obtained from the preceding considerations. For this, we remark that ${\rm
res}L^{n+ {1\over 2}}(x)$ corresponds to
$$
{\rm res}_{\sqrt{\la}=\infty} 2 \la^{n+1} d\sqrt{\la} <1 | n_{+}P_{+}
n_{+}^{-1} | 2 >
$$

$$ = {\rm res}_{\sqrt{\la}=\infty} d\sqrt{\la} < 1 | n_{+}\la^{n+1}
\pmatrix{& {1\over \sqrt{\la}}\cr \sqrt{\la} & \cr} n_{+}^{-1}| 2 >\ ,
$$
and so
$$
\partial_{m} {\rm res} L^{n+{1\over 2}}(x)\qqbox{to} {\rm
res}_{\sqrt{\la}=\infty} \sqrt{\la} d\sqrt{\la} < 1 | [ ( n_{+}a_{-m}
n_{+}^{-1})_{+} \ , n_{+} a_{-n} n_{+}^{-1}]| 2 >
$$
(recall that $a_{-n}$ is represented by the matrix $\pmatrix{0 & \la^{n}\cr
\la^{n+1} & 0\cr}$). Since we have \break ${\rm res}_{\sqrt{\la}=\infty}
\sqrt{\la} d
\sqrt{\la} < 1 | [ x, y]| 2> =0$ if both $x$ and $y$ belong to $\widehat n_{+}
=\CC
f \oplus \la^{-1} s\ell_{2}[[ \la^{-1} ]]$ or to $\widehat b_{-} = \CC e
\oplus \CC
h \oplus \la s\ell_{2}[\la]$ the last expression is symmetric in $n$ and $m$.
(Note the formal analogy between this argument and the classical one.)

\medskip
\noindent
{\bf Acknowledgements~:} I would like to thank B.~Feigin, A.~Orlov and
V.~Rubstov for useful and stimulating discussions on the themes of this paper,
and also the organizers of the Alushta conference for their kind invitation.

\medskip
\noindent
{\bf References~:}

\bibitem{[Di]} L.A. Dickey, Soliton Equations and Hamiltonian Systems, Advanced
Series in Mathematical Physics, vol.12, World Sci., Singapore, 1992.
\medskip
\bibitem{[DS]} V.G. Drinfeld, V.V. Sokolov, Jour. Sov. Math., 30 (1985),
1975-2036.
\medskip
\bibitem{[FF]} B. Feigin, E. Frenkel, Integrals of motions and quantum groups,
hep-th 93100222.

\adresse{Centre de math\'{e}matiques\cr
URA 169 du CNRS\cr
Ecole Polytechnique\cr
91128 Palaiseau Cedex\cr}


\bye